# Using Relational Problems to Teach Property-Based Testing

John Wrenn[a], Tim Nelson[a], and Shriram Krishnamurthi[a]

a   Brown University, USA

**Abstract**

**Context**   The success of QuickCheck has led to the development of *property-based testing* (PBT) libraries for many languages and the process is getting increasing attention. However, unlike regular testing, PBT is not widespread in collegiate curricula.

Furthermore, the value of PBT is not limited to software testing. The growing use of formal methods in industry, and the growth of software synthesis, all create demand for techniques to train students and developers in the art of specification writing. We posit that PBT forms a strong bridge between testing and the act of specification: it's a form of testing where the tester is actually writing abstract specifications.

**Inquiry**   Even well-informed technologists mention the difficulty of finding good motivating examples for its use. We take steps to fill this lacuna.

**Approach & Knowledge**   We find that the use of "relational" problems – those for which an input may admit multiple valid outputs – easily motivates the use of PBT. We also notice that such problems are readily available in the computer science pantheon of problems (e.g., many graph and sorting algorithms). We have been using these for some years now to teach PBT in collegiate courses.

**Grounding**   In this paper, we describe the problems we use and report on students' completion of them. We believe the problems overcome some of the motivation issues described above. We also show that students can do quite well at PBT for these problems, suggesting that the topic is well within their reach. In the process, we introduce a simple method to evaluate the accuracy of their specifications, and use it to characterize their common mistakes.

**Importance**   Based on our findings, we believe that relational problems are an underutilized motivating example for PBT. We hope this paper initiates a catalog of such problems for educators (and developers) to use, and also provides a concrete (though by no means exclusive) method to analyze the quality of PBT.

# The Art, Science, and Engineering of Programming

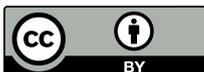

 **Introduction**

*Property-based testing* (PBT) has gone from a research idea [6] to an increasingly-accepted practice [23] in industry in a relatively short period.[1] As systems get more complex, PBT becomes increasingly useful in checking system quality. However, PBT is barely discussed in software engineering texts; is not part of standard computing curricula like the ACM/IEEE's [20]; and, anecdotally, does not appear to be part of many computer science programs (especially outside research universities). There are many things needed to introduce a new practice into curricula (such as time), but adoption also depends on quality problems that can motivate it for students.

In parallel, recent years have seen the growing use of formal methods in industry [2, 4, 5, 9, 14, 26]. Their utility creates the demand for techniques to train students and developers in their use. This need will only grow with the increasing power of tools such as synthesizers. Many of these tools – verifiers, synthesizers, etc. – depend on the ability to write specifications.

We posit that PBT is useful here too, serving as a bridge between programming and specification. PBT has the virtue that it is a form of testing, an activity whose industrial importance students readily appreciate (in our anecdotal experience, even more so if they have industrial experience, e.g., through an internship). At the same time, PBT fundamentally depends on being able to write abstract assertions that are effectively specifications. Thus, we consider PBT a form of "Trojan horse" for introducing students to a specification mindset.

However, little has been written on how to motivate the use of PBT. Indeed, a recent Twitter discussion initiated by Hillel Wayne, an industrial advocate of and consultant in formal methods, is telling [30]. Wayne asks for reasons why people don't use PBT, and while some answers include problems caused by non-determinism, even well-informed technologists mention the difficulty of finding good motivating examples [11] for its use. An earlier (and more caustic) discussion by David R. MacIver [22] focuses specifically on the need for good motivating examples.

In contrast, we have been using PBT in collegiate courses for several years now with what we believe is some success. We work with students who are somewhat familiar with unit testing (but not necessarily any more testing than that). We then point out that when a unit test fails, there are usually two explanations. The most common is that the function under test is wrong; less commonly, the test is at fault.[2] In some cases, however, neither is to blame. Rather, the problem might be *relational* in nature: i.e., a given input might have multiple valid outputs. For such problems, it is natural – and when the number of possible outputs is large, essentially unavoidable – to write a property or specification that characterizes the input-output relationship. Even quite standard and early problems, such as sorting or graph algorithms, are relational in their generality.

---

[1] Even if the reader feels there is insufficient *current* acceptance, surely they can accept that there is value in preparing our educational system for the future.

[2] A third possibility involves impurity, but the examples we use in our classes, and in this paper, are all pure.

In this paper, we describe the problems we use and the course contexts where we have tested them. We also devise a useful method to evaluate the accuracy of student-authored specifications and characterize their common mistakes. We apply this to the student work to find that, largely, they do quite well at PBT.

Therefore, we believe relational problems are an effective way to motivate PBT. They address the lacuna identified by Wayne and others. We do not claim they are universally sufficient or the only kinds of problems students should be shown. However, we believe that through the use of such problems, we can bring PBT to a broad range of programmers, thereby also preparing them to use tools that depend on formal specification.

In short, our thesis is: *relational problems are an underused source of PBT motivation; many problems in the canonical CS pantheon are already relational; and college students can PBT these problems well*.

## 2   Course Structures and Content

Our case study is set at Brown, a highly selective, private US university. We assess the property-based tests of undergraduate students in two courses.

**CS-AccInt**    CS-AccInt is an accelerated introductory course on algorithmic concepts up through basic graph algorithms and big-O analysis, along with functional techniques like streams. It is taught in Pyret, an ML-flavored functional language with algebraic datatypes and an optional static type system (which students are not required to use, though a few do). Over 90% of the population is in their first semester of college (18 years old). Students can only gain entry into the course by successfully doing assignments from several chapters (through recursive list processing and basic higher-order functions) of *How to Design Programs* [12], using Racket. Most students have some prior computing experience, but few with functional programming and essentially none with Pyret. Most of the course is strictly functional. There were 64–88 students (number differs by year) in the course.

CS-AccInt did not have much explicit classroom instruction about PBT. Rather, the process was explained through a few paragraphs in the first PBT assignment, accompanied by a brief discussion by the professor (Krishnamurthi) in class. The class has a strong emphasis on unit-testing (which is graded), with students additionally using Examplar [32] to check their understanding of the problem statement through (non-PBT) tests, so the broader context of testing did not need to be introduced.

**CS-AppLog**    CS-AppLog is a developer-oriented (rather than primarily theoretical) undergraduate course on applied logic that focuses on concrete tools such as Alloy [17], Spin [16], Isabelle [27], and Dafny [21]. The only prerequisite is completing introductory computer science (such as CS-AccInt), so students range from first-year undergrads to graduate students, though second- and third-year undergrads are most numerous. There were 92–147 students (number differs by year) in the course.

CS-AppLog students received one lecture on pbt before the Toposortacle (section 3.3) assignment was released. The lecture motivated pbt and gave students practice writing an oracle for a purported sorting implementation. (In short, this was much more direct classroom instruction than in CS-AccInt.) It also briefly summarized topological sorting, since some students might not have seen it before.

## 3    Key Assignments

Between CS-AccInt and CS-AppLog, three assignments introduce students to the idea of relational problems and explicitly task them with developing a testing infrastructure for them. In each assignment, students implement the predicate oracle, which consumes a purported implementation of a specified function and is satisfied if that implementation is apparently correct.[3] As sub-tasks of implementing oracle, students are required to implement two additional functions, with these shapes of types:

1. generate-input :: Number → Output
   Given a size $n$, produces a random input of size $n$ for the tested function.

2. is-valid :: (Input, Output) → bool
   Satisfied if the property relating the tested function's Input to its Output holds.

To implement oracle, students compose generate-input and is-valid for various input sizes, and additionally test the purported implementation against a handful of small, manually-crafted inputs. To assess how well students reasoned about properties, we focus solely on students' is-valids.

We chose not to assess students' oracle functions, as that would partly also evaluate their ability to generate inputs (both via generate-input and with hand-crafted cases) for the tested function. We do not assess students' generate-input functions in this work, as we did not teach advanced input generation strategies. (Students were asked to implement it to get familiar with the idea of input generation, which is useful in other contexts as well, such as fuzzing – which, after all, is also a kind of pbt.)

The source of a correct implementation of the tested function was available to students for Sortacle and Matcher. In contrast, for Toposortacle, students were provided pseudocode for topological sort along with an executable; students were asked to implement the pseudocode to build their intuition (many had not seen topological sort before). We now briefly describe each assignment. The original text of the assignments is provided in our supplementary material.

---

[3] We say "apparently" because, naturally, testing cannot detect all failures in these problems, which have unbounded inputs. For readability, we drop this style of caveat in the rest of this paper.

### 3.1 Sortacle[4] (CS-AccInt)

Students implement a testing oracle for functions specified to consume a list of people, and produce that list sorted in ascending order by age. They define an is-valid conforming to this signature:

$$\text{is-valid} :: \big(\text{List}\langle\text{Person}\rangle, \text{List}\langle\text{Person}\rangle\big) \rightarrow \text{bool}$$

A Person structure consists of a name and age. Crucially, Sortacle does *not* dictate that the tested function is a stable sort.

Sortacle is the third assignment of CS-AccInt, released seven days after the start of the semester. Students have five days to complete the assignment. By this point students have had significant practice with list-recursive functions over one or two lists, including tasks that are arguably harder than sorting (such as very basic document similarity and collaborative filtering).

### 3.2 Matcher[5] (CS-AccInt)

Students implement a testing oracle for functions solving the stable marriage problem of matching companies to candidates. They define an is-valid conforming to this signature:

$$\text{is-valid} :: \big(\text{List}\langle\text{List}\langle\text{Number}\rangle\rangle, \text{List}\langle\text{List}\langle\text{Number}\rangle\rangle, \text{Set}\langle(\text{Number}, \text{Number})\rangle\big) \rightarrow \text{bool}$$

The lists-of-lists encode the ranked preferences of candidates and companies; the set of pairs encodes a stable matching.

Matcher is the fifth assignment of CS-AccInt, released fourteen days after the start of the semester. Students have five days to complete the assignment. In the one assignment between Sortacle and Matcher, students write several more data-processing mini-tasks inspired by Fisler, Krishnamurthi, and Siegmund [13].

Matcher challenges students to apply PBT to a problem they most probably cannot solve. At the time Matcher is assigned, the course has not introduced the necessary algorithmic techniques, and we expressly do not expect students to solve it. Recall that the assignment provides students with the source code of a correct implementation, which can use refer to in addition to using it to test their oracle.

### 3.3 Toposortacle[6] (CS-AppLog)

Students implement both a topological sorter and a testing oracle for such sorters. They define an is-valid conforming to this signature:

$$\text{is-valid} :: \big(\text{List}\langle\text{Number}, \text{Number}\rangle, \text{List}\langle\text{Number}\rangle\big) \rightarrow \text{bool}$$

---

[4] https://cs19.cs.brown.edu/{2017, 2018, 2019}/sortaclesortacle.html
[5] https://cs19.cs.brown.edu/{2017, 2018, 2019}/oracleoracle.html
[6] https://cs.brown.edu/courses/cs195y/{2018, 2019}/historical/oracle.pdf

Toposortacle is the first assignment of CS-AppLog, assigned two days after the start of the semester. Students have nine days to complete the assignment. Students are free to use either Python or Java to complete the assignment?[7]

The work in this course challenges students to determine and encode properties in (nearly universally) unfamiliar formal methods tools. Therefore, Toposortacle invites them to begin this process in the familiar environment of programming, starting with something they know (testing) but giving it a twist (properties) to set them up for thinking abstractly about problems.

## 4 Experiences

How can students' is-valids inform us about their problem misconceptions? Given that is-valid is a binary classifier of Input–Output pairs, we *could* test each is-valid on a procedurally-generated set of Input–Output pairs, and assess its accuracy with the classical measures of precision and recall. These measures might quantify how buggy students' is-valids are, but it provides no *semantic* insight into their shortcomings.

Instead, we decompose the property embodied by each is-valid into an equivalent conjunction of several sub-properties, and develop, for each property, a test suite which accepts is-valids enforcing that property and rejects those that do not.[8] The particular decomposition we select for each problem corresponds, roughly, to the granularity at which this paper's authors reason about the programming sub-tasks of that problem in their pedagogical context. (Different sub-property decompositions from the ones we selected might be more informative for different pedagogical contexts.) Students were, however, not shown these sub-properties, nor asked to use them to decompose their predicates.

We will develop the details of this method through a concrete example, Sortacle (section 4.1). This description effectively generates a template for assessment, which we also instantiate on Matcher (section 4.2) and Toposortacle (section 4.3).

### 4.1 Sortacle

To evaluate whether novice students can PBT a problem for which they could probably devise a solution, we assess submissions of Sortacle from CS-AccInt's 2017, 2018, and 2019 offerings.

---

[7] In past years students were free to use any language of their choice, but knowledge of I/O operations – to deal with the data format – caused students to converge to these two languages, especially Python, which are now explicitly supported. We will have more to say about language choice in sections 5.2 and 6.

[8] As noted earlier, this classification is inherently imperfect: a test suite can demonstrate the presence of bugs for *particular* inputs, but not the absence of bugs for *all* inputs. Our use of the terms *accepts* and *rejects* should be interpreted with this caveat in mind.

**Sub-Properties**   We decompose sortedness into three sub-properties relating a list (lst) with its sorted self (srt):

1. SAME-SIZE: the sizes of lst and srt must be the same.
2. SAME-ELEMENTS: the members of lst and srt must be the same.
3. ORDERED: the elements of srt must be ordered correctly.

**Sub-Property Suites**   We concretize each of these sub-properties with a distinct test suite that will *reject* is-valids that fail to uphold that particular property:

1. ENFORCE-SAME-SIZE rejects is-valid if it produces true for an input with lst and srt are different sizes, but nonetheless contains the SAME-ELEMENTS and are ORDERED.
2. ENFORCE-SAME-ELEMENTS rejects is-valid if it produces true for inputs where the sets of elements in lst and srt differ, but nonetheless are the SAME-SIZE and ORDERED.
3. ENFORCE-ORDERED rejects is-valid if it produces true for an input where srt is unordered, but nonetheless contains the SAME-ELEMENTS and is the SAME-SIZE as lst.

Each of the three sub-property suites use inputs that violate only their respective property. For instance, consider our ENFORCE-SAME-SIZE suite:

```
1  check "Enforce-Same-Size":
2    a = person('A', 42)
3    len-4 = [list: a, a, a, a]
4    len-7 = [list: a, a, a, a, a, a, a]
5
6    is-valid(len-4, len-7) is false
7    is-valid(len-7, len-4) is false
8  end
```

The inputs len-4 and len-7 contain the SAME-ELEMENTS and are ORDERED, but are *not* the SAME-SIZE.

**The RELATIONAL Suite**   As students have a correct sorter available to them, a tempting path of least resistance is to define is-valid simply as sort(lst) == srt, where sort is the instructor-provided sorting function. This is-valid will behave correctly for inputs where lst admits only one valid ordering, but not for inputs where lst admits *multiple* valid orderings (e.g., an input where two people have the same names but different ages). To detect such faulty is-valids, we construct a test suite dubbed "RELATIONAL". In RELATIONAL, we construct an input lst that admits multiple valid orderings, and test is-valid with *every* possible srt admitted by lst. RELATIONAL rejects is-valid if it produces false for any of these inputs.

**The FUNCTIONAL Suite**   An is-valid that always produces false will be accepted by the sub-property suites, but clearly does not exemplify any of their progenitorial sub-properties. Likewise, an is-valid that always produces true is accepted by RELATIONAL, but this hardly demonstrates the student respected the relational aspect of the problem. Such is-valids are not merely a hypothetical concern: for various reasons (such as insufficient time to complete the assignment), students occasionally submit deeply

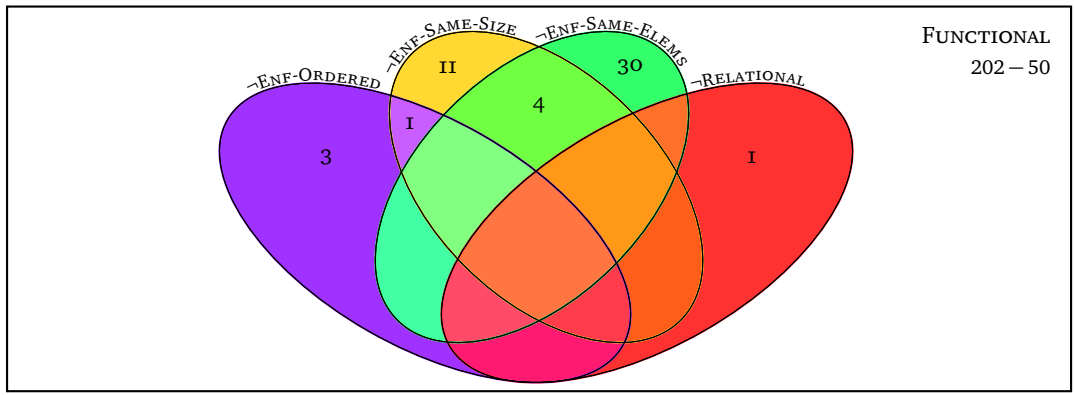

■ **Figure 1** In this Venn diagram, the areas of intersection reflect particular patterns of acceptance and rejections. For instance, 4 is-valids were accepted by ENFORCE-ORDERED and RELATIONAL, but rejected by ENFORCE-SAME-SIZE and ENFORCE-SAME-ELEMENTS. The unlabeled regions are unpopulated.

incomplete is-valids. To identify the is-valids with conceptually interesting flaws, we must first identify these 'trivially' buggy is-valids. We do this with a test suite, dubbed "FUNCTIONAL", that accepts an is-valid only if it produces both:

- true for a srt that is the *only* valid value of the given lst
- false for a srt and lst that do not satisfy *any* of the enumerated properties.

We use this FUNCTIONAL suite to filter out the 'uninteresting' failures of the other suites (ENFORCE-SAME-SIZE, ENFORCE-SAME-ELEMENTS, ENFORCE-ORDERED, and RELATIONAL), by only examining the is-valids that are first accepted by FUNCTIONAL.

**Analysis** We applied our mechanism to evaluate 205 CS-ACCINT Sortacle submissions produced in 2017, 2018, and 2019. Of these submissions, only 3 were not FUNCTIONAL. Among the remaining 202 submissions, only 50 failed any of our subproperty suites:

- 4 students were rejected by ENFORCE-ORDERED.
- 16 students were rejected by ENFORCE-SAME-SIZE.
- 34 students were rejected by ENFORCE-SAME-ELEMENTS.
- 1 student was rejected by RELATIONAL.

Figure 1 illustrates the number of students with each acceptance–rejection pattern. (The top right shows the universe size, 202, minus the 50 members of the inner sets.)

That only 3 is-valids were not FUNCTIONAL should *not* be interpreted to mean that all but 3 lacked simple bugs. For instance, 16 is-valids behaved incorrectly (usually causing exceptions) if either argument was an empty list.[9] We purposely avoided test cases using such small inputs in our other suites, so that these test suites might better reflect how students thought about the properties writ large, rather than how they coped with edge-case inputs. Had we included tests of input-size edges in the

---

[9] We know this number because the final grading of SORTACLE *did* include tests of input-size edge-cases.

Functional suite, the set of submissions for which we analyzed subproperty-suite failures (i.e., submissions accepted by the Functional suite) would have contained 16 fewer submissions.

It is particularly interesting to drill into the rejections of Same-Elements even further, to gain finer insight into students' errors. We can decompose Enforce-Same-Elements as a conjunction of three suites:

- Enforce-Retain rejects is-valid if it produces true for an input where lst is not a subset of srt,
- Enforce-No-New rejects is-valid if it produces true for an input where srt is not a subset of lst,
- Enforce-Not-Disjoint rejects is-valid if it produces true for an input where lst shares no elements in common with srt.

Among the is-valids rejected by Enforce-Same-Elements,

- 6 were rejected by only Enforce-Retain,
- 11 were rejected by only Enforce-No-New,
- 1 was accepted by only Enforce-Not-Disjoint.

These 18 'directional' failures account for slightly more than half of Enforce-Same-Elements's aforementioned 34 rejections. The remaining rejections are accounted for by additional variants of these three suites constructed using inputs with people differing only in one respect (e.g., name, but not age).

## 4.2 Matcher

Recall that we study Matcher as an example of a problem for which we find it unlikely students could envision an implementation.

**Sub-Properties**    We identify three sub-properties relating is-valid's Input (the preferences of companies and candidates) to its Output (a purported match).

1. Stable: there are no two members represented in the match, one from each set, that would prefer each other to their current match.
2. Unique: each member represented in the match is paired with exactly one member of the other set.
3. Complete: all members in the Input preference lists are represented in match.

**Sub-Property Suites**    We concretize each of these sub-properties with a test suite:

1. Enforce-Stable accepts is-valids that produce false for inputs where two non-matched members in match would prefer each other, but are nonetheless Unique and Complete.
2. Enforce-Unique accepts is-valids that produce false for inputs where a member in the match is paired with more than one member of the other set, but are nonetheless Stable and Complete.

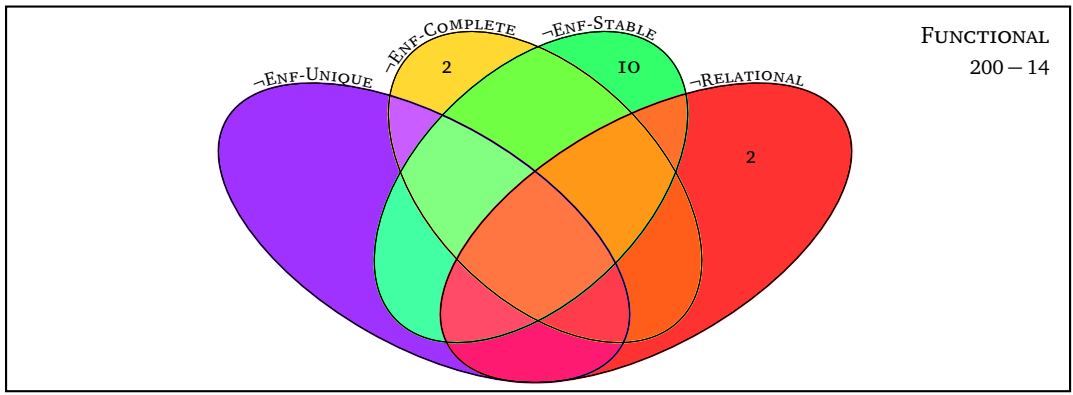

**■ Figure 2**   In this Venn diagram, we see that all students failing any of Matcher's sub-property suites failed *exactly* one sub-property suite. Unlabeled regions reflect unpopulated rejection patterns.

3. Enforce-Complete accepts is-valids that produce false on inputs where a member in either preference list is not represented in the match, but are nonetheless Stable and Unique.

As before, we develop a Functional suite, to detect trivially broken is-valids, and a Relational suite, to detect is-valids that did not respect that a set of preferences could admit multiple stable matches.

We initially struggled to conceive of inputs for Matcher that would violate precisely one property. The types of Input and Output, and their relationship to each other, are more complex than was the case with Sortacle. For Matcher, we ultimately employed a model-finder to produce suitable inputs. We constructed an Alloy* [24] model in which each of our three properties was realized as a predicate. Then, for each property, we requested instances of the model for which that predicate was violated, but the others were satisfied.

**Analysis**   We applied our mechanism to evaluate 200 CS-AccInt Matcher submissions produced in 2017, 2018, and 2019. Of these submissions, none was rejected by Functional. Only 14 failed any of our subproperty suites:

- 0 students were rejected by Enforce-Unique.
- 2 students were rejected by Enforce-Complete.
- 10 students were rejected by Enforce-Stable.
- 2 students were rejected by Relational.

Figure 2 illustrates the number of students with each pattern of acceptances and rejections.

### 4.3 Toposortacle

Finally, we assess CS-AppLog submissions for Toposortacle.

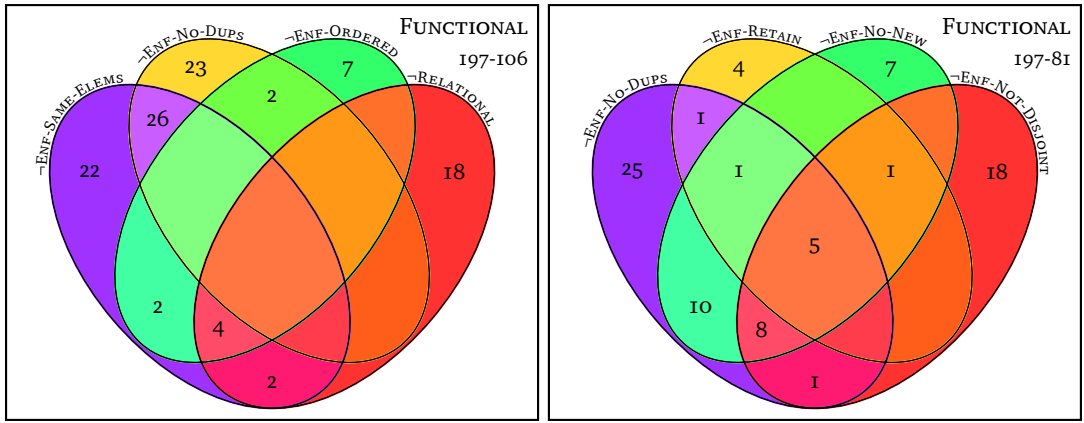

■ **Figure 3** Acceptance–rejection patterns for Toposortacle. At left, the patterns for our four high-level subproperties. At right, the patterns of acceptance–rejection between Enforce-No-Dups and our decomposition of Enforce-Same-Elements.

**Sub-Properties**    We identify three sub-properties relating is-valid's Input (a dag) to its Output (srt):

1. Same-Elements: the sets of vertices present in dag and srt are equal.

2. Ordered: the elements of srt are ordered with respect to the partial order given by dag.

3. No-Dups: srt contains no duplicate vertices.

**Sub-Property Suites**    As before, we concretize each of these sub-properties with a test suite:

1. Enforce-Same-Elements rejects is-valid if it produces true for inputs where the vertices contained in dag and srt differ.

2. Enforce-Ordered rejects is-valid if it produces true for an input where srt is unordered.

3. Enforce-No-Dups rejects is-valid if it produces true for an input where srt contains a vertex more than once.

As before, we additionally develop a Functional suite and a Relational suite.

**Analysis**    We applied our mechanism to evaluate 237 CS-AppLog Toposortacle submissions produced in 2018 and 2019. Of these submissions, 40 were not Functional. Among the remaining 197 submissions, 106 failed any of our subproperty suites:

- 51 students were rejected by Enforce-No-Dups.
- 56 students were rejected by Enforce-Same-Elements.
- 15 students were rejected by Enforce-Ordered.
- 24 students were rejected by Relational.

Seventy students failed exactly one of the subproperty suites, 32 failed exactly two, 4 failed exactly three, and none failed all four. Enforce-No-Dups and Enforce-Same-Elements dominate the observed rejections, and seem to interact: 26 is-valids were rejected *only* by Enforce-No-Dups and Enforce-Same-Elements. We can drill into this interaction by decomposing Same-Elements into a conjunction of sub-suites:

- Enforce-Retain: all vertices of srt are present in dag.
- Enforce-No-New: all vertices of dag are present in srt.
- Enforce-Not-Disjoint rejects is-valid if it produces true for an input where srt and dag contain the same number of vertices, but of completely disjoint identities.

From experience, we expect that a significant number of students will enforce Same-Elements by comparing the number of vertices appearing in dag against the length of the srt, rather than examining the actual vertex identities. This antipattern is detected by Enforce-Not-Disjoint.

Figure 3 illustrates the numbers of students with each acceptance–rejection pattern of our four high-level suites, and the patterns just considering Enforce-No-Dups and our decomposition of Enforce-Same-Elements. As with Sortacle, we find that failure to guard against completely new elements in the output (Enforce-No-New) was more common than failure to guard against the omission of elements (Enforce-Retain). However, *both* of these failure modes were dwarfed by the antipattern of relying on length to ensure Same-Elements (Enforce-Not-Disjoint).

**Examining the Performance Difference**    Because students did less well on Toposortacle than on the other assignments, we now delve further into the data and possible causes.

Of the 40 submissions that did not pass Functional for Toposortacle, 14 were lacking in implementation: either their code was unfinished or threw some exception (section 5.2) on our tests. In contrast, the other 26 actually returned the wrong boolean on some test in Functional. However, of the students who passed Functional, nearly half (46 %, or 91 students) received *perfect* scores, and the vast majority (over 80 %, 91+70=161 students) missed no more than one subproperty – demonstrating learning, albeit imperfectly in some cases.

Nevertheless, the outcome is not as positive as for the other two problems. What might explain this?

We have already noted the difference in student profile (section 2). This assignment was also conducted at a university with a strong "shopping period" culture: for the first two weeks of class, students attend (and do assignments for) many more courses than they will eventually take. Thus, students generally have less time to complete assignments during shopping period. CS-AppLog students worked on Toposortacle during the *first* of these two weeks – when their distractions were at a maximum. Indeed, many of them are unsure if they will take CS-AppLog, and some do not stay in it. (Though Sortacle was also presented early in CS-AccInt, having gone through an admission process, most of those students stay.)

Of the students whose submissions failed Functional, 6 dropped the course, all of whom returned the wrong boolean on some test. The direction of causality is unclear (Toposortacle may have *caused* them to drop!), but it means only 20 students in total returned the wrong boolean on a Functional test and remained in the class.

There is one more, subtler, problem. Toposortacle's input generator is far more complex than the generators for Sortacle and Matcher: students must produce directed acyclic graphs as input, which requires them to reason structurally about graphs rather than just populating a list with random values. Experience at office hours leads us to believe that students spend most of their time on the generator, which distracts from the core purpose of the assignment.

## 5 Discussion

### 5.1 What Have We Learned?

We set out to investigate how students (starting from first-year) do on property-based testing problems. We felt PBT might be too ambitious a task (especially for first-year students) and were worried we might see high error rates or even just incomprehension. Happily, that does not seem to have been the case. While the results on Toposortacle are definitely much more ambiguous, we also see various confounding factors, which we have discussed in section 4.3. We will discuss the assignment more in section 5.4, but we still feel comfortable concluding that, at this level, students seem capable of writing specifications. (Their subsequent success in CS-AppLog also bears this out.)

We noticed in both Sortacle and Toposortacle that students suffered from a "directionality" problem: students' PBT failed to notice new elements introduced in the output that were not in the input. This is reasonable from an operational perspective: while a buggy implementation could easily duplicate an input element in the output, it would be a rare implementation whose output contained entirely new elements that were not in the input. In a specification setting, however, this is a dangerous detail to overlook, because it would (for instance) give a synthesis algorithm freedom to generate an implementation that the specifier almost certainly did not want. Therefore, while this phenomenon is not widespread in *this* student population on *these* problems, its occurrence is still noteworthy, and the transition from PBT to full-on specification would need to address this.[10]

The authors were surprised to find that performance on Matcher was not notably worse than on Sortacle or Toposortacle, given the seemingly much greater complexity and unfamiliarity of the problem. Some of this may have been mitigated by giving students a solution (and a worthwhile research question is the impact of giving them the *source* versus just a black-box implementation). Still, the authors felt Matcher was a more complex problem and expected it to be borne out with diminished accuracy of is-valids, but it was not. This suggests our intuitions about both problem complexity and the mapping from that to testing difficulty may need more calibration. (As a conjecture, perhaps the more complex the problem, the easier it is to find something wrong? Another possibility, which our Reviewer 1 also noted: Perhaps with more complex problems it's harder for the student to envision an implementation. This forces them to think harder about the problem abstractly, which lends itself better to writing a validity checker.)

---

[10] We note that a more abstract formulation of these problems also addresses this – e.g., specifying sorting as permutation, rather than as a collection of sub-properties – but perhaps at some cost to concreteness. Determining which style of these specifications is better for what tasks, and the cognitive effects of each style, is outside the scope of this paper.

## 5.2 Language Errors

Students completing CS-AppLog's Toposortacle assignment could use *either* Java or Python. Of the 212 students who used Python (vs. 25 in Java), 58 produced is-valids that threw exceptions (KeyError and ValueError) on our test suites. In contrast, *none* of the Java is-valids yielded exceptions. Based on manual examination of student code, these exceptions appear to largely arise from two classes of mistake: indexing errors and type-related errors.

Index-related mistakes come from indexing into a dictionary with an invalid key. These most often came from students whose check for SAME-ELEMENTS was one-sided: they would, for instance, loop through the output vertices seeking a matching input vertex in an auxiliary dictionary. When our test suites gave output vertices that didn't appear in the input, their programs had no safety check in place.

Type-related mistakes occur where students seem to have misunderstood our specification or made unsafe assumptions. For instance, one student expected *numbers* for vertex identifiers, rather than *strings*, and another student anticipated entries in the input to consist of *lists* rather than *tuples*. It is interesting that these problems occurred only in the non-statically-typed language.

## 5.3 What Sub-Property Decomposition Misses

In this paper, we assess assignments through the lens of just one rubric: sub-property decomposition. However, there are several things this approach misses. For concreteness, we analyze this in more detail in the context of Sortacle.

For one, it only reveals what students do wrong – a strict subset of the space of things students do that are pedagogically interesting! For instance, in the preceding analysis, we reflected only on the properties characterizing a valid sort, but what constitutes a valid Person is also constrained by the assignment: their age shall be non-negative. Although the specification of is-valid does not suggest that a lst containing negative ages cannot be sorted, some students nonetheless relied on this constraint in their is-valids: of the 202 CS-AccInt Sortacle is-valids accepted by FUNCTIONAL, 27 rejected otherwise valid inputs that contained people with negative ages. This behavior was not the consequence of students actively forbidding inputs with invalid people, but rather a side-effect of a coding shortcut: assuming that the minimum age of an empty list was some negative number.

The sub-property decomposition approach is also poorly suited to detecting certain kinds of programming errors. For instance, the Sortacle specification does not dictate an upper bound on Person ages. Despite this, 6 of the FUNCTIONAL is-valids rejected (otherwise valid) inputs containing 'unreasonably' old people. How old is unreasonably old? From manual inspection of the is-valids, we find: 100 years, 123 years,[11] 1000 years (from two students), 1001 years, and 9 999 999 999 years of age. As before, this behavior did not stem from students actively forbidding inputs with old people, but rather a side-effect of a coding shortcut: assuming that the maximum age of an empty list was some unreasonably large number. With one grim exception (who discriminated against centenarians), these students correctly (with respect to the real world – *not* the specification) superimposed their own understanding of the constraints of human longevity.

### 5.4 Evolving Toposortacle

As we have seen, student performance in Toposortacle was less than exemplary. We have identified some possible explanations, such as the different student population, the structure of "shopping period", and the complexity of the generator.

The first of these is not in our control. The second, working around shopping period, technically is. However, our goal of using PBT as a gateway to formal specification precludes moving the assignment later.

Nevertheless, we can address the issue of input generation. One potential fix would be to ask them to write is-valid for the DAG generator itself, and provide pseudocode for that – rather than introducing the excess cognitive load of topological sorting. We are also considering changing the allowed-languages list to avoid type-related errors (Section 5.2).

### 5.5 Generation Versus Checking

Our assignments expect students to both generate random inputs (generate-input) and to check implementations (is-valid). For the narrow purpose of this paper, only is-valid matters, and having them write the former might be seen as a distraction, or even a problem (as we note in section 5.4).

From a broader pedagogic perspective, however, we believe there is still some value to writing generate-input. First, it serves as a useful exercise in recursive programming. More subtly, students are traditionally more accustomed to writing functions whose outputs are not much bigger than their input, but generate-input turns this around radically. Second, we believe it demystifies generate-input. Rather than viewing a sophisticated input generator as a magical black box, we believe students are more likely to view it merely as a better version of something they can already write, and perhaps to even critically investigate how it is better. If they end up in a language that does not already have a robust PBT framework, they know how to build one from scratch, rather than giving up on the idea entirely. Nevertheless, a course more focused on PBT could investigate these issues in some depth, perhaps also seeing whether there is any (causal) relationship between the quality of input generator and quality of validity predicate.

---

[11] This is one year older than the dying age of Jeanne Calment, considered by many [8] to be the oldest person to have ever lived.

Therefore, the question of whether or not students should write generate-input must be viewed in the broader pedagogic context. This should perhaps also be combined with explicit instruction on how to write a quality input generator – something we have not had time to do in our courses. In situations where it becomes an active problem (section 4.3), it should perhaps be modified (section 5.4) or removed entirely.

## 5.6 Other Examples of PBT

The motivation for this paper is to find instances of PBT that are accessible to even early college students. Our experience, reported in this paper, is that relational problems work well for this purpose, where the entire focus of the assignment is PBT. There are many more problems that are arguably just as compelling for using PBT. Concretely, in CS-AccInt, we find PBT was useful to students on several exercises assigned after Sortacle and Matcher (which is why those assignments went out early in the semester):

- Students implement and test continued fractions using lazy streams. Because streams are represented with thunks, they cannot be directly compared for equality; students must define property-based predicates to test their streams.
- Students implement and test an alternate, append-based representation of lists [29]. The function that splits a list into two parts is not guaranteed to behave deterministically, forcing them to use PBT.
- Students implement and test Dijkstra's algorithm; a graph may admit multiple equally-short paths between two vertices.
- Students implement and test Prim's and Kruskal's minimum spanning tree algorithms; a graph may admit multiple minimum spanning trees.
- Students implement and test seam-carving [3] using both memoization and dynamic programming; an image can admit multiple eliminable seams.

In some cases the assignments allude to the importance of using PBT, while others explicitly tell them to do so. Furthermore, other tool support in the course (we depend on Wrenn and Krishnamurthi's Examplar tool [32], which was originally designed to help students understand the problem definition through tests) also makes them realize the need for PBT.

This constellation of problems hopefully helps students appreciate that PBT is not a niche technique but rather one of broad applicability across several kinds of problems and several areas of computer science. Our use of PBT in CS-AppLog suggests that PBT provides a pathway to training students in specification – an increasingly marketable skill. Anecdotally, we have received feedback from former students who report that their job interviewers were impressed with their testing skills. We hope these experiences embolden other computer science educators to adopt PBT as a foundational skill in their courses.

## 6    From Art to Science

Until now, we have only reported on observations from our teaching experience. These are useful for obtaining an early understanding and formulating hypotheses. We now briefly discuss some of the ways this work could inspire research.

**What do run-time errors teach us?**    We are intrigued by the stark differences in programming language errors observed between Python and Java programmers on Toposortacle. Similarly, we observe that drastically more CS-AppLog students failed the Toposortacle's Functional suite than CS-AccInt students did (programming in Pyret) on Sortacle and Matcher. Programming language choice may therefore have some impact on students' pbt experiences.

Small sample size aside, the data do not allow us to arrive at a causal argument. In particular, assuming the phenomenon scales up, there are still many possible explanations for what we see:

1. They could be inherent to the languages (e.g., the more forgiving overloading in the semantics of Python might encourage sloppier programming).
2. They could be cognitive effects generated by the styles of the languages (e.g., Java's static typing may cause programmers to think through their data more carefully, even when the specific problems might not be caught by the language's type system).
3. They could be reflections on the programmers (e.g., perhaps more careful programmers choose Java, more sloppy ones choose Python).
4. They could be driven by affordances of the libraries, which combines the first two points above (more permissive libraries cause programmers to take more liberties, which eventually end up hurting their programs).

Research methods that suggest causal explanations would be necessary to tease apart these phenomena. Readers may find useful some recent surveys [7, 25], about human-factors methods, some of which can help make causal conclusions.

**Are we missing testing errors?**    We believe our subproperty-based methodology is an interesting first step toward analyzing student performance, but it also has several weaknesses. Section 5.3 already discusses how our methodology misses certain kinds of pedagogically-interesting mistakes. A different analysis – based, for instance, on test cases that exhibit violations of combinations of properties – would lead to a more fine-grained understanding of student work. However, these are not always easy to create: as noted, we had to use a model-finder to generate some of our tests. A more thorough, formally-grounded methodology would enable us to make more comprehensive statements about student performance, which is a prerequisite to many kinds of subsequent research.

**How do testing failures map to implementation failures?**    We have an underlying assumption that particular patterns of rejection and acceptance correspond to particular buggy is-valid implementation strategies. For instance, we expect that students with similar acceptance–rejection patterns have similar conceptual bugs. However, this is only a hypothesis; we would need to perform a careful inspection of the is-valids (manual or automated) to be able to draw a firm conclusion. Performing this analysis is important, because in its absence, instructors do not get enough insight (or, worse, might get misleading "insight") into what their students are doing well or poorly.

**What is the student experience?**   We have anecdotal evidence that students accept PBT and even appreciate its value in helping them find positions (section 5.6). An interesting research question would ask how student understanding of testing, PBT, and specification evolves over the course of their exposure to these problems. It is common in education to not only examine outcomes but also study the evolution of students' self-image and perceptions of ability; in this setting, there are additional characteristics, such as students' sense of software as something whose reliability is questionable and must therefore be thoroughly vetted. While our experience suggests we will find growth in these areas, carefully designing studies and evaluating their responses remains open.

**How does student background impact performance?**   We note that students do rather worse in Toposortacle than the other two assignments. CS-AppLog also has a broader cross-section of the student population. Some students in the class are much more familiar with functional programming, and even with PBT, than others. What impact does this have? A reliable conclusion about the feasibility of PBT needs to be indexed by the background of students.

## 7   Related Work

Our literature review found relatively little on teaching PBT. We present related work here in order of increasing relevance.

Fredlund, Herranz, and Mariño [15] teach a course on concurrency where students specify pre- and post-conditions on robot behavior. The course uses property-based testing to evaluate projects. Here the use of property-based testing mostly appears beneficial to instructors, rather than being deliberately presented to students.

In Dadeau and Tissot's course [10], students write abstract models in the B notation [1]. The key innovation is that, rather than the traditional B approach of gradually refining the model until an implementation is reached, students use their initial models to generate tests via simulation. Students are then challenged to detect program mutants using their own generated tests. While model-based testing provides an intellectual connection between our approaches, ours (asking students to encode properties programmatically and test randomly generated inputs) is quite different.

Scharff and Sok [28] describe adding a program-correctness module to a programming languages course. The module begins with Design-by-Contract style pre- and post-conditions (enforced by assertions and exceptions). It then graduates to proving correctness for the same exercises that students previously wrote contracts for. It would be possible to adapt the correctness predicates that our students write to postcondition assertions, but our focus is not only on postconditions as a precursor to Hoare-style correctness proofs, but rather on property-based testing more broadly.

A recent book [31] by Wikström illustrates PBT using several interesting scenarios from a screencast editor. While a fascinating case study, most examples require interest in and knowledge of manipulating video; they are also generally more sophisticated than ours. As a result, that work may not fit into most curricular contexts, whereas we deal with topics widely covered in computing curricula. Of course, for those immersed in video, Wikström's examples are probably far superior to ours.

Jahangirova, Clark, Harman, and Tonella [18, 19] show how to use techniques such as code mutation and automatic test generation to assist developers in validating and improving oracles. They evaluate this approach empirically with both students and professional programmers, and find it to be effective. However, their work focuses on the improvement of existing oracles, represented as assert statements in existing code. In contrast, we are interested in how to best teach PBT to students.

**Acknowledgements**  This work is partially supported by the US National Science Foundation. We thank the anonymous reviewers, especially Reviewer 1, for their careful reading and detailed comments. We are also grateful to Hillel Wayne for discussions on this topic.

## About the authors

**John Wrenn** (jswrenn@cs.brown.edu) is a guerilla archivist, tandem bicycle evangelist, and PhD student at Brown University.

**Tim Nelson** (tbn@cs.brown.edu) preaches the good news of logic and computing at Brown University.

**Shriram Krishnamurthi** (sk@cs.brown.edu) is the Vice President of Programming Languages (no, not really) at Brown University.